\documentclass[twoside,a4paper]{article}
\usepackage{amsmath,graphicx,amssymb,fancyhdr,amsthm}
\newtheorem{thm}{Theorem}[section]
\newtheorem{cor}[thm]{Corollary}
\newtheorem{lem}[thm]{Lemma}
\newtheorem{prop}[thm]{Proposition}

\theoremstyle{definition}
\newtheorem{defn}[thm]{Definition}
\theoremstyle{remark}

\def\beq{\begin{eqnarray}}
\def\eeq{\end{eqnarray}}
\def\bsp{\begin{split}}
\def\esp{\end{split}}

\def\d{\mathrm{d}}

\def\RC{$CSI_R$}
\def\KC{$CSI_K$}
\def\FC{$CSI_F$}
\newcommand{\FCSI}[1]{$CSI_{F,#1}$}
\newcommand{\mf}[1]{{\mathfrak #1}}

\newcommand{\mbold}[1]{\mbox{\boldmath{\ensuremath{#1}}}}
\newcommand{\scalar}[2]{\left\langle{#1},{#2}\right\rangle}

\begin{document}

\title{\Large\textbf{Lorentzian spacetimes with constant curvature invariants in four dimensions}}
\author{{\large\textbf{Alan Coley$^{1}$, Sigbj\o rn Hervik$^{2,~ 1}$, Nicos Pelavas$^{1}$} }
 \vspace{0.3cm} \\
$^{1}$Department of Mathematics and Statistics,\\
Dalhousie University,
Halifax, Nova Scotia,\\
Canada B3H 3J5
\vspace{0.2cm}\\
$^{2}$Faculty of Science and Technology,\\
 University of Stavanger,\\  N-4036 Stavanger, Norway
\vspace{0.3cm} \\
\texttt{aac, pelavas@mathstat.dal.ca, sigbjorn.hervik@uis.no} }
\date{\today}
\maketitle
\pagestyle{fancy}
\fancyhead{} 
\fancyhead[EC]{A. Coley, S. Hervik and N. Pelavas}
\fancyhead[EL,OR]{\thepage}
\fancyhead[OC]{Lorentzian $CSI$ spacetimes in four dimensions}
\fancyfoot{} 

\begin{abstract}

In this paper we investigate four dimensional Lorentzian
spacetimes with constant curvature invariants ($CSI$ spacetimes).
We prove that if a four dimensional spacetime is $CSI$, then
either the spacetime is locally homogeneous or the spacetime is a
Kundt spacetime for which there exists a frame such that the
positive boost weight components of all curvature tensors vanish
and the boost weight zero components are all constant. We discuss
some of the properties of the Kundt-$CSI$ spacetimes and their
applications. In particular, we discuss $\mathcal{I}$-symmetric
spaces and degenerate Kundt $CSI$ spacetimes.

\end{abstract}

\newpage

\section{Introduction}

Lorentzian spacetimes for which all polynomial scalar invariants
constructed from the Riemann tensor and its covariant derivatives
are constant ($CSI$ spacetimes) were studied in \cite{CSI} and
\cite{CSI3d}.
In this paper we will investigate $CSI$ spacetimes in four dimensions
(4D) utilizing a number of the results
derived in \cite{chpinv}.

The definitions and notation used in this paper will follow those of
\cite{CSI,CSI3d} (and a summary of the algebraic classification of tensors and 
and a summary of curvature operators,
necessary for this paper, will be given in the Appendices). In particular,
we consider a spacetime $\mathcal{M}$ equipped with a metric $g$ and let $\mathcal{I}_k$
denote the set of all polynomial scalar
invariants constructed from the curvature tensor and its covariant derivatives up
to order $k$. 

\begin{defn}[{$VSI$}$_k$ spacetimes]
$\mathcal{M}$ is called {$VSI$}$_k$ if for any invariant $I\in\mathcal{I}_k$, $I=0$ over $\mathcal{M}$.
\end{defn}
\begin{defn}[{$CSI$}$_k$ spacetimes]
$\mathcal{M}$ is called {$CSI$}$_k$ if for any invariant $I\in\mathcal{I}_k$, $I$ is constant (i.e., $\partial_{\mu}{I}=0$) over $\mathcal{M}$. 
\end{defn}
Moreover, if a spacetime is {$VSI$}$_k$ or {$CSI$}$_k$ for all
$k$, we will simply call the spacetime {$VSI$} or {$CSI$},
respectively. The set of all locally homogeneous spacetimes will be
denoted by $H$. Clearly $VSI \subset CSI$ and $H \subset CSI$.

\begin{defn}[{\sc \RC} spacetimes]
Let us denote by \RC ~
all reducible $CSI$ spacetimes that can be built from $VSI$ and
$H$ by (i) warped products (ii) fibered products, and (iii) tensor sums.
\end{defn}

\begin{defn}[{\sc \FC} spacetimes]
Let us denote by \FC~ 
those spacetimes for which there exists a frame with a null vector
$\ell$ such that all components of the Riemann tensor and its
covariants derivatives in this frame have the property that (i)
all positive boost weight components (with respect to $\ell$) are zero and (ii) all
zero boost weight components are constant.
\end{defn}
Note that \RC$ \subset CSI$ and \FC$
\subset CSI$. (There are similar definitions for \FCSI{k}
etc.).

\begin{defn}[{\KC} spacetimes]
Finally, let us denote by \KC, those $CSI$
spacetimes that belong to the (higher-dimensional) Kundt class (defined later);
the so-called Kundt $CSI$ spacetimes.
\end{defn}


We recall that a spacetime is Kundt on an open neighborhood if it
admits a null vector $\ell$ which is geodesic, non-expanding,
shear-free and non-twisting; i.e.
\[\ell^{\mu}\ell_{\nu;\mu}=0, \quad \ell^{\mu}_{~;\mu}=\ell^{\mu;\nu}\ell_{(\mu;\nu)}=\ell^{\mu;\nu}\ell_{[\mu;\nu]}=0 \]
(which leads to constraints on the
Ricci rotation coefficients in that neighborhood; namely the
relevant Ricci rotation coefficients are zero). We note that the Kundt
$CSI$ spacetimes are actually degenerate Kundt spacetimes, in
which there exists a common null frame in which the geodesic,
expansion-free, shear-free and twist-free null vector $\ell$ is
also the null vector in which all positive boost weight terms of
the Riemann tensor and all of its covariant derivatives are zero
\cite{chpinv}. We also note that if the Ricci rotation
coefficients are all constants then we have a locally homogeneous
spacetime.

In addition, we will make extensive use of the algebraic classification of
the Riemann tensor (i.e., Riemann type) and its covariant
derivatives according to boost weight, as  described in detail in \cite{coley} (also see 
Appendix A, where the algebraic types of the Riemann and its covariant derivative, $\nabla$Riemann, are given).
In particular, in 4D the resulting Weyl type is equivalent to the Petrov type,
and the Ricci type is related to the PP-type (which is related to the
Segre type of the Ricci tensor). We will primarily use the terms Petrov type
and Segre type\footnote{Here, and later, Segre type will refer to the Segre type of the Ricci tensor unless otherwise stated.} here to be consistent with the usual terminology in 4D, and only
use the terms Weyl type and Ricci type when referring to results of interest in higher
dimensions.

For a Riemannian manifold, there is no distinction between the $CSI$ requirement and locally homogeneity: every $CSI$ is locally
homogeneous $(CSI \equiv H)$ \cite{PTV}. This is not true for Lorentzian
manifolds.  However, for every $CSI$ spacetime with particular constant
invariants there is a homogeneous spacetime (not necessarily
unique) with precisely the same constant invariants. From the work in \cite{CSI}
it was conjectured
that if a spacetime is {$CSI$} then the Riemann tensor (and hence the
Weyl tensor) is of type {\bf II}, {\bf III}, {\bf N} or {\bf O} \cite{coley}
(part of the ${CSI_F}$ conjecture), that if a spacetime
is {$CSI$}, the spacetime is either locally homogeneous or
belongs to the higher dimensional Kundt $CSI$ class (the ${CSI_K}$
conjecture) and that it can be constructed from locally
homogeneous spaces and vanishing scalar invariant ($VSI$)
spacetimes \cite{4DVSI} (the ${CSI_R}$
conjecture). The various $CSI$ conjectures were proven in three dimensions
in \cite{CSI3d}.

There are a number of important applications of $CSI$ spacetimes.
First, there are examples of  $CSI$ spacetimes
which are exact solutions in supergravity and hence of particular
physical interest \cite{CFH}, including  $AdS \times S$ spacetimes
\cite{Freund}, generalizations of $AdS \times S$ based on
different $VSI$ seeds \cite{CFH}, and the $AdS$ gyratons
\cite{frolov}. $CSI$ spacetimes are also
related to universal spacetimes \cite{CGHP}.

Second, the characterization of $CSI$ spacetimes is useful for
investigating the question of when a spacetime can be uniquely
characterized by its curvature invariants. This question was
addressed in \cite{chpinv}, where the class of four dimensional
Lorentzian manifolds that can be completely characterized by the
scalar polynomial curvature invariants constructed from the
Riemann tensor and its covariant derivatives was determined. In
particular, an appropriate set of projection operators  was derived from the
Riemann tensor and its covariant derivatives which
enabled the proof of a number of results, including the main
theorem that a metric that is not characterized by its curvature
invariants must be of Kundt form.

\section{Main theorems}
The main theorem is the following:

\begin{thm}
A 4D spacetime is $CSI$ if and only if either:
\begin{enumerate}
\item{} the spacetime is locally homogeneous; \emph{or} \item{}
the spacetime is a \emph{Kundt spacetime} for which there exists a
frame such that all curvature tensors have the following
properties: (i) all positive boost weight components vanish; (ii)
all boost weight zero components are constants.
\end{enumerate}
\end{thm}
\begin{proof}
The first part of this theorem follows from the corollary  in
\cite{chpinv}, which says that either a $CSI$ metric is locally
homogeneous or it is Kundt.

The second part of this theorem, which essentially says that \KC
~implies \FC, also follows from the work of \cite{chpinv} by keeping
track of the boost weight 0 components of the curvature tensors.
The most effective way of seing that all the boost weight 0
components can be put into a constant form is to utilize the
curvature operators.  For example, the Petrov type II or D case
gives curvature projectors, $\bot_1$ and $\bot_2$, of type
$\{(1,1)(11)\}$ already at the level of the Weyl tensor (see Appendix B
for the notation and a short review of curvature operators).  These
projectors are already aligned with the Kundt frame so we 
immediately obtain that the boost weight 0 components of the Weyl
tensor are constants.  The idea is that the curvature projector
picks out the necessary boost weight 0 components and relates them
to invariants (possibly in an algebraic way).  The components are
thus constants provided the invariants are constants.  By doing
this for the full Riemann tensor and its derivatives it becomes
clear that for a \KC ~spacetime, in the Kundt frame, all boost
weight 0 components are determined by the curvature invariants in
the sense of \cite{chpinv}.  Hence, if the invariants are constants,
the boost weight 0 components must also be constants.  \end{proof}

{\it Note that all of the $CSI$ conjectures are
proven in 4D as a result of this theorem.}

\begin{thm}
For a 4D Lorentzian spacetime,
\[ CSI \Leftrightarrow CSI_3.\]
\end{thm}
\begin{proof}
The result that $CSI\Rightarrow CSI_3$ is trivial. Assume,
therefore, that the spacetime is $CSI_3$.

From the calculations in \cite{chpinv} we see that as we reach the
third order invariants, by inspection of all the cases, we either
have constructed our timelike projector $\bot_1$ or not.  For the
cases where we have the projector, the metric is
$\mathcal{I}$-non-degenerate, and hence we have a spacetime which is
$CH_3$, where $CH_k$ denotes curvature homogeneous of order $k$
\footnote{Recall that a $k$-curvature homogeneous spacetime is
defined as a spacetime for which there exists a frame such that the
components of the curvature tensors up to order $k$ are all constants.}.
In \cite{Mp} the relation between curvature homogeneity and
locally homogeneity was studied;  In particular, it follows that
$CH_3$ implies local homogeneity and consequently $CSI$.

If we do not have a timelike projector, the spacetime is either
Kundt or it is Petrov type $O$, Segre type $\{(1,11)1\}$. Treating
the latter case first, the $CSI_0$ assumption and the
Bianchi identity immediately imply that $\nabla(Riem)=0$ (and thus the
spacetime is symmetric). Hence, this is a locally homogeneous
space and thus $CSI$.

Lastly, assume the spacetime is Kundt. This part of the proof will
be given in section \ref{sect:Kundt}.
\end{proof}

\section{Kundt $CSI$ metrics}\label{sect:Kundt}
From the above and using the results of \cite{CSI}, we have now
established that a Kundt $CSI$ spacetime can be written in the form \beq \d
s^2=2\d u\left[\d v +H(v,u,x^k)\d u+W_{i}(v,u,x^k)\d
x^i\right]+g^{\perp}_{ij}(x^k)\d x^i\d x^j, \label{HKundt} \eeq
where $\d S^2_H=g^{\perp}_{ij}(x^k)\d x^i\d x^j$ is the locally homogeneous
metric of the `transverse' space and
the metric functions $H$ and $W_ {i}$, requiring $CSI_0$,
are given by \beq
W_{i}(v,u,x^k)&=& v{W}_{i}^{(1)}(u,x^k)+{W}_{i}^{(0)}(u,x^k),\label{HKa}\\
H(v,u,x^k)&=& {v^2}\tilde{\sigma}+v{H}^{(1)}(u,x^k)+{H}^{(0)}(u,x^k), \label{HKb} \\
\tilde{\sigma} &=& \frac 18\left(4\sigma+W^{(1)i}W^{(1)}_i\right),
\label{sigma}
\eeq
where $\sigma$ is a constant.
The remaining equations for $CSI_0$ that need to be solved are (hatted indices refer to an orthonormal frame in the transverse space):
\beq
 \label{Wcsi1}
W^{(1)}_{[\hat i;\hat j]} &=& {\sf a}_{\hat i\hat j}, \\
W^{(1)}_{(\hat i;\hat j)}-\frac 12 \left(W^{(1)}_{\hat
i}\right)\left(W^{(1)}_{\hat j}\right) &=& {\sf s}_{\hat i\hat j},
\label{Wcsi4}\eeq and the components ${R}^{\perp}_{\hat i\hat
j\hat m\hat n}$ are all constants (i.e., $\d S^2_H$ is curvature homogeneous). Here the 
antisymmetric and symmetric constant matrices ${\sf a}_{\hat i\hat j}$ and 
${\sf s}_{\hat i\hat j}$, respectively, are determined from the boost weight 0 components of the Riemann tensor (see eqs. (\ref{Defas}-\ref{DefasEnd}) later).
In four dimensions, $\d S^2_H$ is 2-dimensional, which immediately implies  $g^{\perp}_{ij}(x^k)\d
x^i\d x^j$ is a 2-dimensional locally homogeneous space and is, in
fact, a maximally symmetric space.  Up to scaling, there are
(locally) only 3 such spaces; namely, the sphere, $S^2$, the flat
plane, $\mathbb{E}^2$, and the hyperbolic plane, $\mathbb{H}^2$.

The equations (\ref{Wcsi1}) and (\ref{Wcsi4}) now give a set of
differential equations for $W^{(1)}_{\hat i}$. These equations
uniquely determine $W^{(1)}_{\hat i}$ up to initial conditions
(which may be free functions in $u$).

Requiring also $CSI_1$, gives a set of constraints:
\beq
{\mbold\alpha}_{\hat i}&=&\sigma W^{(1)}_{\hat i}-\frac 12({\sf s}_{\hat j\hat i}+{\sf a}_{\hat j\hat i})W^{(1)\hat j}, \\
{\mbold\beta}_{\hat i\hat j\hat k}&=&W^{(1)\hat n}{R}^{\perp}_{\hat n\hat i\hat j\hat k}-W^{(1)}_{\hat i}{\sf a}_{\hat j\hat k}+({\sf s}_{\hat i[\hat j}+{\sf a}_{\hat i[\hat j})W^{(1)}_{\hat k]},
\eeq
where ${\mbold\alpha}_{\hat i}$ and ${\mbold\beta}_{\hat i\hat j\hat k}$ are constants determined from the boost weight 0 components of the covariant derivative of the Riemann tensor (see eqs. (60-63) in \cite{CSI}).

\begin{lem} For a 4D Kundt spacetime, $CSI_1$ implies $CSI$.
\label{lemCSI}\end{lem}
\begin{proof}
To show this we therefore need to show that the above equations
are sufficient to ensure that the spacetime is $CSI$. The $CSI_0$
and $CSI_1$ conditions force the boost weight 0 components of $Riem$
and $\nabla(Riem)$ to be constants.

The boost weight 0 components of Riemann are
\beq \label{Defas}
R_{0101}=-\sigma, \quad R_{01ij}={\sf a}_{ij}, \\
R_{0i1j}=\frac 12({\sf s}_{ij}+{\sf a}_{ij})\\
R_{ijkl}=c(\delta_{ik}\delta_{jl}-\delta_{jk}\delta_{il}). 
\label{DefasEnd}
\eeq
These components split into boost weight 0
components of the Weyl and Ricci tensors. From these we can
construct curvature operators in the way described in \cite{chpinv} and Appendix B. Combining both the
Weyl operator and the Ricci operator we get projectors of the
following types, which we consider in turn. In the following the Segre-like notation refers to the set of curvature projectors \cite{chpinv}. 
\begin{itemize}
\item{} $\{(1,1)11\}$: This is the general case and it implies
that $({\sf s}_{ij}+{\sf a}_{ij})$ has two distinct eigenvalues.
Further, from the $CSI_1$ criterion, we see that $W^{(1)}_i$ is an
eigenvector of this matrix, unless it is a constant. If it is a
constant, we immediately get $CSI$. Therefore,  assume it is an
eigenvector. Again we get that either it is a constant, in which
case this implies $CSI$, or it is not constant and
${\mbold\alpha}_i=0$. We also note that we also must have
${\mbold\beta}_{ijk}=0$ and, in fact, $(\nabla Riem)_0=0$. Therefore,
since we cannot acquire higher boost weight components by taking
the covariant derivatives (due to the fact it is Kundt), this also
implies $CSI$. \item{} $\{(1,1)(11)\}$: By considering $(\nabla
Riem)_0$, we get two possibilities. If $(\nabla Riem)_0\neq 0$, we can
always construct a second spacelike projector so that we have a
set $\{(1,1)11\}$. This would again imply that $W^{(1)}_i$ is a
constant; hence, this spacetime is $CSI$. If $(\nabla Riem)_0= 0$,
also get a $CSI$ spacetime (for the same reason as above). \item{}
$\{(1,11)1\}$: This case must have $(C)_0=0$. Therefore, the Ricci
tensor is the only tensor with boost weight 0 components. By
calculating $\nabla(Riem)$, we get $(\nabla Riem)_0=0$, and hence this is
a $CSI$ spacetime. \item{} $\{(1,111)\}$: This again gives
$(\nabla Riem)_0=0$, and hence this is a $CSI$ spacetime.
\end{itemize}
\end{proof}

Since we have restricted ourselves to four dimensions,
$g^{\perp}_{ij}(x^k)\d x^i\d x^j$ is a 2-dimensional locally
homogeneous space. Up to scaling, there are (locally) only the
sphere, $S^2$, the flat plane, $\mathbb{E}^2$, and the hyperbolic
plane, $\mathbb{H}^2$. To delineate all the cases we need to integrate the abovementioned equations in terms of the remaining functions
${W}_{i}^{(1)}(u,x^k)$. To solve for these we need to consider these cases in turn.

Below we have listed the various cases in terms of the transverse metric $\d s^2_{\perp}=g^{\perp}_{ij}(x^k)\d x^i\d x^j$, the Segre type of the Ricci tensor, and the one-form ${\bf W}^{(1)}\equiv W^{(1)}_i\d x^i$. The constant $\sigma$ is a freely specifiable constant, unless stated otherwise.

\subsection{The Sphere, $S^2$} \label{sect:sphere}
\paragraph{Segre type $\{2(11)\}$ or $\{(1,1)(11)\}$}
Here,
\beq
 \d s^2_{\perp}=\d x^2+\frac{1}{\lambda^2}\sin^2(\lambda x)\d y^2, \quad {\bf W}^{(1)}=0.
\eeq
The possible Petrov types are D or II (possibly III, N, or O in degenerate cases).
\paragraph{Segre type $\{(211)\}$, $\{(31)\}$, or $\{(1,111)\}$}
Here, we have several cases. They all have $\sigma>0$, and are as follows:
\[ \d s^2_{\perp}=\d x^2+\frac{1}{\sigma}\sin^2(\sqrt{\sigma} x)\d y^2,\]
where
\begin{enumerate}
\item{} ${\bf W}^{(1)}=2\sqrt{\sigma}\tan(\sqrt{\sigma}x)\d x$.
\item{} ${\bf W}^{(1)}=2\sqrt{\sigma}\left[-\cot(\sqrt{\sigma}x)\d x+\tan(\sqrt{\sigma}y)\d y\right]$.
\item{} ${\bf W}^{(1)}=2\sqrt{\sigma}\left[-\cot(\sqrt{\sigma}x)\d x-\cot(\sqrt{\sigma}y)\d y\right]$.
\end{enumerate}
The possible Petrov types are III, N, or O.
\subsection{The Euclidean plane, $\mathbb{E}^2$} \label{sect:flat}
Here we can set
\[ \d s^2_{\perp}=\d x^2+\d y^2.\]
\paragraph{Segre type $\{211\}$ or  $\{(1,1)11\}$}
\begin{enumerate}
\item{}  ${\bf W}^{(1)}=2r\d x, \quad 0\neq \sigma\neq -r^2\neq 0.$
\end{enumerate}
The possible Petrov types are D or II.
\paragraph{Segre type $\{(21)1\}$, $\{31\}$ or  $\{(1,11)1\}$}
\begin{enumerate}
\item{} $\sigma>0$:
${\bf W}^{(1)}=2\sqrt{\sigma}\tan\left(\sqrt{\sigma}x\right)\d x$.
\item{} $\sigma<0$:
${\bf W}^{(1)}=-2\sqrt{|\sigma|}\tanh\left(\sqrt{|\sigma|}x\right)\d x$.
\item{} $\sigma<0$:
${\bf W}^{(1)}=2\sqrt{|\sigma|}\coth\left(\sqrt{|\sigma|}x\right)\d x$.
\item{} $\sigma<0$: ${\bf W}^{(1)}=2\sqrt{|\sigma|}\d x$.
\end{enumerate}
Here, the possible Petrov types are III, N, and O.
\paragraph{Segre type $\{2(11)\}$ or $\{(1,1)(11)\}$}
\begin{enumerate}
\item{} $\sigma\neq 0$: ${\bf W}^{(1)}=0$.
\end{enumerate}
Possible Petrov types are D and II.
\paragraph{Segre type $\{(211)\}$, $\{(31)\}$, or $\{(1,111)\}$}
\begin{enumerate}
\item{} $\sigma=0$: ${\bf W}^{(1)}=\frac{2\epsilon}{x}\d x$, where $\epsilon=0,1$.
This is the $VSI$ case so the possible Petrov types are III, N and O.
\end{enumerate}
\subsection{The Hyperbolic plane, $\mathbb{H}^2$}  \label{sect:hyper}

\paragraph{Segre type $\{211\}$ or  $\{(1,1)11\}$}
\begin{enumerate}
\item{} $\d s^2_{\perp}=\d x^2+e^{-2qx}\d y^2, \quad {\bf W}^{(1)}=\alpha\d x+ \beta e^{-qx}\d y$.
\end{enumerate}
The possible Petrov types are D and II (or III, N in degenerate cases).
\paragraph{Segre type $\{(21)1\}$, $\{31\}$ or  $\{(1,11)1\}$}
\begin{enumerate}
\item{} $ \d s^2_{\perp}=\d x^2+e^{-2qx}\d y^2, \quad {\bf W}^{(1)}=\alpha\d x+ \beta e^{-qx}\d y$, where
\[\sigma=-q^2\pm\tfrac 14\sqrt{(\alpha^2+\beta^2)[\beta^2+(\alpha-2q)^2]}.\]
\end{enumerate}
The possible Petrov types are D and II (or III, N in degenerate cases).
\paragraph{Segre type $\{2(11)\}$ or $\{(1,1)(11)\}$}
\begin{enumerate}
\item{}  $\d s^2_{\perp}=\d x^2+e^{-2qx}\d y^2, \quad {\bf W}^{(1)}=2q\d x, \quad \sigma\neq -q^2$.
\item{}$\d s^2_{\perp}=\d x^2+e^{-2qx}\d y^2, \quad {\bf W}^{(1)}=0, \quad \sigma\neq -q^2$.
\end{enumerate}
The possible Petrov types are D and II (or III, N, O in degenerate cases).
\paragraph{Segre type $\{(211)\}$, $\{(31)\}$, or $\{(1,111)\}$} For all of these $\sigma<0$ and we set $\sigma=-q^2$.
\begin{enumerate}
\item{}  $\d s^2_{\perp}=\d x^2+e^{-2qx}\d y^2, \quad {\bf W}^{(1)}=2q\d x+\frac{2\epsilon}{y}\d y$, where $\epsilon=0,1$.
\item{}$\d s^2_{\perp}=\d x^2+e^{-2qx}\d y^2, \quad {\bf W}^{(1)}=0$.
\item{} $\d s^2_{\perp}=\d x^2+\frac 1{q^2}\sinh^2(qx)\d y^2, \quad {\bf W}^{(1)}=-2q\tanh(qx)\d x$.
\item{} $\d s^2_{\perp}=\d x^2+\frac 1{q^2}\sinh^2(qx)\d y^2, \quad {\bf W}^{(1)}=2q\left[\coth(qx)\d x-\tanh(qy)\d y\right]$.
\item{} $\d s^2_{\perp}=\d x^2+\frac 1{q^2}\cosh^2(qx)\d y^2, \quad {\bf W}^{(1)}=2q\coth(qx)\d x$.
\item{} $\d s^2_{\perp}=\d x^2+\frac 1{q^2}\cosh^2(qx)\d y^2, \quad {\bf W}^{(1)}=2q\left[-\tanh(qx)\d x+\coth(qy)\d y\right]$.
\end{enumerate}
For case 2 the Petrov types can be II or D, while all other cases
have Petrov types III, N or O.


\section{4D locally homogeneous Lorentzian spaces}

From Theorem 2.1 and the discussion in the Introduction, it
follows that every 4D $CSI$ spacetime is either locally homogeneous
or Kundt. In the last section we presented all of the $CSI$-Kundt
spacetime metrics. Let us next briefly discuss the possible 4D
locally homogeneous spacetimes.

A homogeneous space can be considered as a coset manifold $\mathcal{M}=G/H$,
where $G$ is a Lie group and $H$ is a closed Lie subgroup of $G$, equipped
with an invariant metric.  For a homogeneous space the geometry is determined
locally at a point, and consequently to a homogeneous space, $\mathcal{M}=G/H$,
we can associate a homogeneous triple $(\mf{g},\mf{h},\scalar{-}{-})$, where
$\mf{g}$ is the Lie algebra of $G$, $\mf{h}\subset\mf{g}$, and $\scalar{-}{-}$
is a scalar product (the restriction of the invariant metric $g$ at a point)
on a vector space complement $\mf{m}$ such that $\mf{g}=\mf{h}\oplus\mf{m}$.
In the Lorentzian case, the metric $g$ is a Lorentzian metric and this
requires that $\mf{h}\subset\mf{s}\mf{o}(n-1,1)$, where $n=\dim{\mathcal{M}}$.
The classification of locally homogeneous Lorentzian spaces can thus be
reduced to finding such homogeneous triples up to isometry.

In four dimensions, the classification of such triples up to isometry is not
complete.  However, there are some partial classification results.
In \cite{Kom95}, Komrakov classifies all homogeneous pairs $(\mf{g},\mf{h})$
with $\dim{\mf{h}}\geq 1$, and in \cite{Kom01} Komrakov provides all invariant
Lorentzian metrics to these homogeneous pairs.  These lists are quite
extensive and, if correct,  in principle gives all locally
homogeneous spaces with non-trivial isotropy group.  Due to the shear number
of these triples the possible geometries they describe have not been fully
explored.  However, the possible solutions to the Einstein-Maxwell equations
were explored in \cite{Kom01}, and the Ricci-flat geometries were considered
in detail in \cite{Hakenberg}.

The 4-dimensional homogeneous triples where $\dim\mf{h}=0$ can be considered
to correspond to homogeneous spaces being a 4-dimensional Lie group equipped
with a left-invariant Lorentzian metric.  The Lie algebras of dimension 4 are
well-known (see, e.g., \cite{PSWZ}); these can be equipped with a left-invariant
metric in a standard manner \cite{Milnor:76}.

An example of a Lie group equipped with a left-invariant metric is the following vacuum Petrov type I metric:
\beq \d s^2&=&-e^{qz}\left(\cos\left[\tfrac{\sqrt{3}}2qz\right]\d x
-\sin\left[\tfrac{\sqrt{3}}2qz\right]\d t\right)^2\nonumber
+e^{qz}\left(\sin\left[\tfrac{\sqrt{3}}2qz\right]\d x
+\cos\left[\tfrac{\sqrt{3}}2qz\right]\d t\right)^2\\ && +e^{-2qz}\d y^2+\d z^2.\nonumber
\eeq

\section{$\mathcal{I}$-symmetric spaces}
Let us introduce the concept of an $\mathcal{I}$-symmetric space,
which is defined as a \emph{spacetime having the same scalar
polynomial curvature invariants as that of a symmetric space.}
This implies that, by considering the scalar invariants only, we
cannot distinguish an $\mathcal{I}$-symmetric space from that of a
symmetric space (similarly, we can think of  $CSI$ spacetimes as
$\mathcal{I}$-homogeneous spaces).

Recall that a symmetric space is defined as the vanishing of the
first covariant derivative of the Riemann tensor; i.e., for a
symmetric space $R_{\alpha\beta\gamma\delta;\mu}=0$.  This implies
that \emph{the only possible non-vanishing invariants of an
$\mathcal{I}$-symmetric space are the zeroth order invariants}. In
addition, all zeroth order invariants are constant (since the
covariant derivative of the Riemann tensor is zero, for any
invariant $I$ constructed from the Riemann tensor it follows that
$\nabla{I}=0$ and hence $I$ is constant). Therefore, these spaces
have a very simple set of invariants, which may be useful in
applications in other theories of gravity.

The \emph{symmetric} spaces in 4D are all classified and are one of the following:
\begin{enumerate}
\item{} Maximally symmetric space (Segre type $\{(1,111)\}$,
Petrov type O) \item{} Product of a 3-dimensional maximally
symmetric space and $\mathbb{R}$ (Segre type $\{(1,11)1\}$ or
$\{1,(111)\}$, Petrov type O) \item{} Product of two 2-spaces of
constant curvature (Segre type $\{(1,1)(11)\}$, Petrov type D or
O) \item{} A Cahen-Wallach (CW) plane wave spacetime (Segre type
$\{(211)\}$, Petrov type N)
\end{enumerate}
An $\mathcal{I}$-symmetric space must therefore have the same scalar invariants as
these symmetric spaces; in particular, it must be $CSI$. Indeed, the $\mathcal{I}$-symmetric spaces are classified as follows:
\begin{prop}
A spacetime is $\mathcal{I}$-symmetric if and only if it belongs to one of the following classes:
\begin{enumerate}
\item{} Symmetric spaces.
\item{} The following Kundt $CSI$ spacetimes described in section \ref{sect:Kundt}:
\begin{enumerate}
\item{} section \ref{sect:sphere}: All.
\item{} section \ref{sect:flat}: Segre types  $\{(21)1\}$, $\{31\}$,$\{(1,1)1\}$,$\{2(11)\}$, $\{(1,1)(11)\}$, $\{(211)\}$, $\{(31)\}$ and $\{(1,111)\}$.
\item{} section \ref{sect:hyper}: Segre types  $\{2(11)\}$ with ${\bf W}^{(1)}=0$, $\{(1,1)(11)\}$ with ${\bf W}^{(1)}=0$, $\{(211)\}$, $\{(31)\}$ and $\{(1,111)\}$.
\end{enumerate}
\end{enumerate}
\end{prop}
The proof of this is straightforward by considering all the cases in section \ref{sect:Kundt} separately.  We note that many examples
from the literature are $\mathcal{I}$-symmetric (for example, the
Siklos spacetime \cite{Siklos}, the AdS gyraton \cite{Caldarelli,FZ} and many more \cite{CSI,CFH}).

Similarly, we can define a $k$th order $\mathcal{I}$-symmetric space as a
spacetime having the same invariants as a $k$-symmetric spacetime.  Recall that
a $k$-symmetric spacetime is defined by the requirement that
$R_{\alpha\beta\gamma\delta;\mu_1\dots\mu_k}=0$.  Defining the set of $k$th
order $\mathcal{I}$-symmetric spacetimes as $\mathcal{I}\mathrm{Sym}^k$, we get
the sequence of inclusions:  \[
VSI\subset\mathcal{I}\mathrm{Sym}^1\subset\cdots\subset\mathcal{I}\mathrm{Sym}^k.\]

In \cite{Senovilla} the class of 2-symmetric spacetimes was investigated.  It was found that a 2-symmetric
spacetime is either a symmetric spacetime or admits a covariantly constant null vector ($CCNV$).
{\footnote{Therefore, a 2-symmetric spacetime is either locally homogeneous (and hence, CSI-Kundt), or
there exists a covariantly
constant null vector ($CCNV$-Kundt) (a similar result has been conjectured for the set of
$k$-symmetric spacetimes).  Note that a 2-symmetric spacetime need not be $CSI$
(which is locally homogeneous or Kundt),
in which case a $CCNV$ exists.  }}
In the first
instance, we refer to the list of symmetric spacetimes given above; therefore, this subclass of 2-symmetric
spacetimes are necessarily $CSI$ (locally homogeneous).  In particular, they have curvature
invariants that are constant at zeroth order and vanishing at first and higher order.  Evidently, in this
case the following classes are equivalent: $\mathcal{I}\mathrm{Sym}^2=\mathcal{I}\mathrm{Sym}^1$.  In the
second instance, the 2-symmetric spacetimes admitting a $CCNV$ define a subclass of the Brinkmann metrics
(which are themselves a subclass of the Kundt metrics)
\begin{equation} ds^2 =
2du\left[dv+H(u,x^{k})du+W_{i}(u,x^{k})dx^{i}\right]+g_{ij}(u,x^{k})dx^{i}dx^{j} \, , \label{brinkmann}
\end{equation} \noindent
where the $CCNV$ is $\ell=\partial_{v}$.
A study of spacetimes admitting a
$CCNV$ shows that, in general, the Riemann tensor (of metric (\ref{brinkmann}))
and all of its covariant derivatives are of
type II or less \cite{mcnutt}.  Moreover,
all boost weight 0 components of $\nabla^{(k)}(Riem)$, $k\geq 0$, arise solely from the
Riemannian metric of the transverse space, and more importantly $\nabla^{(k)}(Riem)\cdot\ell=0$ 
for all
$k\geq 0$.  This last property describes the form of the non-vanishing components
of Riemann and its derivatives in a null tetrad with $\ell$ being one of the null frame 
vectors.  Therefore,
a 2-symmetric $CCNV$ spacetime implies that the transverse space of
(\ref{brinkmann}) is 2-symmetric; thus
from \cite{Senovilla} it follows that $g_{ij}$ is, in fact, a family of Riemannian symmetric spaces
parameterized by $u$. That is, $\nabla(Riem)$ has vanishing boost weight 0 components, $R_{ijkl;m}=0$, resulting in
$\nabla(Riem)$ of type III, N or O (this last algebraic type corresponds to a symmetric $CCNV$ spacetime).
In a 2-symmetric $CCNV$ spacetime, curvature invariants at first and higher order vanish.  In addition, since
the zeroth order invariants are determined only by the transverse space metric, which are Riemannian
symmetric spaces, it follows that {\em at any point} in a 2-symmetric $CCNV$ spacetime there exists a symmetric $CCNV$
spacetime having the same set of constant zeroth order invariants.
However, a 2-symmetric $CCNV$ spacetime need not have constant invariants on a neighborhood
and, in general, it is therefore not symmetric.
In this sense we have $\mathcal{I}\mathrm{Sym}^1 \subseteq
\mathcal{I}\mathrm{Sym}^2$.  Equality only occurs when $u$ is a non-essential
coordinate in the transverse
metric; that is, $u$ does not parameterize inequivalent Riemannian symmetric spaces $g_{ij}$ thus a
coordinate transformation exists such that $g_{ij}(u,x^{k}) \longrightarrow g_{ij}(x^{k})$.

\section{Degenerate Kundt $CSI$ spacetimes}

Non-locally homogeneous $CSI$ spacetimes are contained in the
degenerate Kundt class. A degenerate Kundt spacetime is a
spacetime that admits an aligned null vector which is geodesic,
shear-free, twist-free and expansion-free such that the Riemann
tensor and all of its covariant derivatives are of type II (or more special) in the
same aligned (kinematic) frame \cite{kundt}.
If the Riemann tensor is of type N, III or O in a degenerate Kundt
spacetime, they must be aligned (with the kinematic frame) and from
the theorems of \cite{4DVSI} the spacetime is $VSI$. This implies
that a non-$VSI$ degenerate Kundt spacetime must be of proper Riemann type II
or type D. In particular, a proper (i.e., non-$VSI$) $CSI$ spacetime
is of Riemann type II or D. A degenerate Kundt spacetime is not
$\mathcal{I}$-non-degenerate and hence it is not locally
characterized by its scalar curvature invariants (which are all
constant in the $CSI$ case) \cite{chpinv}.

A $CSI$ spacetime is either degenerate Kundt or locally homogeneous.
Consider a $CSI$ spacetime in which the Riemann tensor and all of
its covariant derivatives are aligned and of algebraic type D
(i.e., of type D to `all orders' or, in short, type D$^k$, see Appendix A).  
Recall that a type D tensor has only boost weight 0 components, and these components consequently
comprise the curvature invariants. Indeed, although
such a spacetime is degenerate Kundt and hence not
$\mathcal{I}$-non-degenerate, it is exceptional in the sense that
it is locally homogeneous.

\begin{thm}
A 4D type D$^k$ $CSI$ spacetime, in which the Riemann tensor $Riem$ and
$\nabla^k(Riem)$ are all simultaneously of type D, is locally
homogeneous.
\end{thm}
\begin{proof}
In \cite{chpinv} it was shown that if there exists a frame in which all of the positive boost weight terms
of the Riemann tensor and its covariant derivatives $\nabla^{(k)} (Riem)$ are zero (in this frame) in 4D,
then it is Kundt\footnote{This can be seen from \cite{chpinv} by inspection of all the cases.}.  This is obviously true in 4D for type D$^k$ (as it is a special case of above).  From
\cite{chpinv} we can deduce that $CSI_0$ implies $CH_0$.  From the proof of Lemma \ref{lemCSI}, it can also
be seen that assuming $CSI_1$ and that $\nabla(Riem)$ is of type D, implies $CH_1$ (since$\nabla(Riem)$ has only boost weight 0 components) .
In fact, at this stage, we already have curvature operators of type $\{(1,1)(11)\}$ or
$\{(1,1)11\}$.  By Singer's theorem \cite{Singer} the first case implies local homogeneity, while for the latter case
we have to consider $\nabla^2(Riem)$.  However, in this case we already have all the curvature operators
necessary to determine all the independent components of $\nabla^2(Riem)$ (since this is also of type $D$).
{\footnote{Note that since this is also of type D the curvature operators obtained from
$\nabla(Riem)$ can be used to determine all boost weight 0 components of $\nabla^{2}(Riem)$
in terms of scalar polynomial invariants up to second order.}}
Hence, $CSI_2$ implies
$CH_2$ and consequently the spacetime is also locally homogeneous.  \end{proof}

These spacetimes, even though they are not $\mathcal{I}$-non-degenerate, are
in some sense `characterized' by their constant curvature
invariants, at least within the class of type D$^k$ $CSI$
spacetimes. In general, there are many degenerate Kundt $CSI$
metrics (that are not type D$^k$) with the same set of constant
invariants,  $\mathcal{I}$. In this case there is at least one
$\nabla^k(Riem)$ which is proper type II and thus has negative
boost weight terms; this Kundt $CSI$ metric will have precisely the
same scalar curvature invariants as the corresponding type D$^k$
$CSI$ metric (which has no negative boost weight terms). Therefore,
there is a distinguished or a `preferred' metric with the same set of
constant invariants, $\mathcal{I}$; namely, the corresponding type D$^k$
locally homogeneous $CSI$ metric, which is distinguished within the
class of algebraic type D$^k$ $CSI$ spacetimes.

It is plausible that the following is true: If $\nabla^{(k)}(Riem)$ is type $D^k$ then every boost
weight 0 component of $\nabla^{(k)}(Riem)$ is, in principle, expressible in terms of $k^{th}$ order
curvature invariants.
As a corollary, if $\nabla^{(k)}(Riem)$ is type $D^k$ and $CSI$ then the spacetime is locally homogeneous.

In proving Theorem 6.1, we note that every Riemann type
D, CH$_2$ spacetime is locally homogeneous \cite{Mp}.  Thus, if a spacetime is type $D^2$ and
$CSI$$_{2}$, then by showing it is CH$_2$ gives the desired result.

It is not hard to show that if the Riemann tensor is of type D, then all boost weight 0 components, $\Psi_2$,
$\Phi_{11}$, $\Phi_{02}$ and $\Lambda$ (using the Newman Penrose (NP) notation) can be expressed in terms of $0^{th}$ order curvature
invariants; hence $CSI$$_0$ implies CH$_0$.  Calculating $\nabla(Riem)$ is now somewhat simpler
since all of these components are just products of constant boost weight 0 curvature scalars and
certain spin coefficients.  When $\nabla(Riem)$ is of type D a refinement of the above statement,
that CH$_2$ implies local homogeneity, can be made by noting the following facts.  Let $G_k$ denote the
isotropy group of $\nabla^{(k)}(Riem)$; then for Riemann type D, $G_0$ is either 2-dimensional
consisting of boost and spins if $\Phi_{02}=0$, or 1-dimensional consisting of only boosts if
$\Phi_{02}\neq 0$.  It was shown in \cite{Mplong} that for the Riemann type D case with
$\Phi_{02}=0$, if $\nabla(Riem)$ is CH$_1$ then $G_1$ cannot be 1-dimensional.  Therefore, if $G_1$ is
2-dimensional, by Singer's Theorem the spacetime is locally homogeneous .  Hence, in this case a
proper CH$_1$ can only occur if $G_1$ is trivial.  In the second case ($\Phi_{02}\neq 0$), we again have
by Singer's Theorem that if $G_1$ is 1-dimensional then the spacetime is locally homogeneous,
implying that a proper CH$_1$ only occurs if $G_1$ is trivial.

In both cases of Riemann type D we find that a proper CH$_1$ (i.e.,  not locally homogeneous) must have (if
it exists) a frame that is completely fixed by $\nabla(Riem)$.  If we suppose that $\nabla(Riem)$ is type D
(in addition to Riemann type D), then since the only non vanishing components are boost weight 0 we always
have, at least, a boost isotropy.  Thus $G_1$ cannot be trivial, implying there is no proper CH$_1$.
Therefore, if a spacetime has Riemann and $\nabla(Riem)$ type D and is CH$_1$ then it is locally
homogeneous.


Note that the order in the above statements can be reduced and
since the proof of Theorem 6.1 shows that CSI$_1$ implies CH$_1$, we have
as a result:
\begin{cor}
Every 4D CSI$_1$ spacetime in which $Riem$ and $\nabla(Riem)$ are
simultaneously of type D is locally homogeneous.
\end{cor}

\noindent
In particular, in 4D if the spacetime is Petrov type D,
then if $\nabla(Riem)$ is type D the spacetime is Kundt.
For Petrov type O, there are several Segre types: for
Segre types $\{(1,1)11\}$ and $\{(1,1)(11)\}$ (PP-type D),  if $\nabla(Riem)$ is type D
the spacetime is Kundt, and
for Segre types $\{(1,11)1\}$ or $\{(1,111)\}$ the spacetime is symmetric and so
$\nabla(Riem)=0$. Therefore,  if the spacetime  is $CSI$ and
$Riem$ and $\nabla(Riem)$ are of type D, it follows that the spacetime is Kundt.

This establishes that spacetimes satisfying the conditions of Corollary 6.2
are members of the Kundt class, and hence this result is already covered by
Lemma 3.1 as a special case when $Riem$ and $\nabla(Riem)$ are of type D.

\section{Discussion}

In this paper we have proven that if a 4D spacetime is $CSI$, then
either the spacetime is locally homogeneous or the spacetime is a
Kundt spacetime. A number of partial results can be deduced from
previous work, some of which are presented in Appendix C. We have
also discussed the properties of the Kundt-$CSI$ spacetimes. The
$CSI$ spacetimes are of particular interest since they are
solutions of supergravity or superstring theory, when supported by
appropriate bosonic fields \cite{CFH}. It is plausible that a wide
class of $CSI$ solutions are exact solutions to string theory
non-perturbatively \cite{string}.

In the context of string theory, it is of considerable interest to
study
higher dimensional Lorentzian $CSI$ spacetimes. In particular,
a number of N-dimensional $CSI$ spacetimes are known to be
solutions of supergravity theory when supported by
appropriate bosonic fields \cite{CFH}.
It is known that $AdS_d \times S^{(N-d)}$ (in short $AdS\times S$)
is an exact solution of supergravity (and preserves the maximal
number of supersymmetries) for certain values of $(d,N)$ and for
particular ratios of the radii of curvature of the two space
forms \cite{Freund}. There are a number of other
$CSI$ spacetimes known to be solutions of supergravity and admit
supersymmetries; namely, generalizations of $AdS \times S$, generalizations of the chiral
null models \cite{hortseyt} and generalizations of the $AdS$ gyraton
\cite{frolov}.

More general supergravity $CSI$ solutions have been constructed by taking
a homogeneous (Einstein) spacetime,
$(\mathcal{M}_{Hom},\tilde{g})$, of Kundt form and generalizing to an
inhomogeneous spacetime, $(\mathcal{M},{g})$,  by including
arbitrary Kundt metric functions \cite{CFH}. In addition,
product manifolds of
the form $M\times K$, where $M$ is an Einstein
space with negative constant curvature and $K$ is a (compact)
Einstein-Sasaki spacetime, can give rise to supergravity
$CSI$ spacetimes.
The supersymmetric properties of $CSI$ spacetimes have also
been studied \cite{mcnutt}. It is known that, in general, if a spacetime admits a
Killing spinor it necessarily admits a null or timelike Killing
vector (KV). Therefore, a necessary (but not sufficient) condition for
a particular supergravity solution to preserve some supersymmetry
is that the spacetime possess such a KV.

In future work, motivated by the physical interest of higher
dimensional $CSI$ spacetimes, we shall discuss possible higher
dimensional generalizations to the results presented in this
paper. In particular, we hope to prove a higher dimensional
version of Theorem 2.1 and to generalize Theorem 6.1 (and thus show
that a Kundt $CSI$ is $CSI$$_F$ in arbitrary dimensions)
\cite{shortinv}. The first step is to investigate the curvature operators in higher dimensions and to classify these for the various algebraic types. With the aid of these operators is it then hoped that the results obtained
in 4D here can also be shown to be true in higher dimensions.

\appendix
\section{Appendix: Algebraic classification}
Given a covariant tensor $T$ with respect to an Newman Penrose (NP) tetrad (or null
frame) $\{\ell, n, m, \bar{m}\}$, the effect of a boost $\ell \mapsto e^{\lambda}\ell$, $n
\mapsto e^{-\lambda}n$ allows $T$ to be decomposed according to
its boost weight,
\begin{equation}
T=\sum_b (T)_{b}, \label{bwdecomp}
\end{equation}
where $(T)_{b}$ denotes the boost weight $b$ components of $T$. Recall that the boost weight $b$ components are defined as those components, $T_{ab...d}$, of $T$ that transform according to
\[ T_{ab...d}\mapsto e^{b\lambda}T_{ab...d},\]
under the aforementioned boost. 

An
algebraic classification of tensors $T$ has been developed
\cite{coley} which is based on the existence of certain
normal forms of (\ref{bwdecomp}) through   successive application
of null rotations and spin-boosts.  In the special case where $T$
is the Weyl tensor in four dimensions, this classification reduces
to the well-known Petrov classification. However, the boost weight
decomposition can be used in the classification of any tensor $T$
in arbitrary dimensions.  As an application, a Riemann tensor of type $G$ 
has the following decomposition,
\begin{equation}
R=(R)_{+2}+(R)_{+1}+(R)_{0}+(R)_{-1}+(R)_{-2},
\end{equation}
\noindent in every null frame.  A Riemann tensor is algebraically special if there 
exists a frame in which certain boost weight components can be transformed to zero; these are summarized in Tables \ref{riemtypes} and \ref{driemtypes}.

\begin{table}[h]
\begin{center}
\small{
\begin{tabular}{c|l}
\hline
 & \\
{\bf Riemann type} & {\bf Conditions}  \\
 & \\
\hline
G & --- \\
I & $(R)_{+2}=0$ \\
II & $(R)_{+2}=(R)_{+1}=0$ \\
III & $(R)_{+2}=(R)_{+1}=(R)_{0}=0$ \\
N & $(R)_{+2}=(R)_{+1}=(R)_{0}=(R)_{-1}=0$ \\
D & $(R)_{+2}=(R)_{+1}=(R)_{-1}=(R)_{-2}=0$ \\
O & all vanish (Minkowski space) \\
\hline
\end{tabular}
}
\caption{The relation between Riemann and the vanishing of boost weight components.  For example, $(R)_{+2}$ corresponds to the frame components $R_{1313},R_{1414},R_{1314}$.}\label{riemtypes}
\end{center}
\end{table} 
\begin{table}[h]
\begin{center}
\small{
\begin{tabular}{c|l}
\hline
 & \\
{\bf $\nabla R$ type} & {\bf Conditions}  \\
 & \\
\hline
G & --- \\
H & $(\nabla R)_{+3}=0$ \\
I & $(\nabla R)_{+3}=(\nabla R)_{+2}=0$ \\
II & $(\nabla R)_{+3}=(\nabla R)_{+2}=(\nabla R)_{+1}=0$ \\
III & $(\nabla R)_{+3}=(\nabla R)_{+2}=(\nabla R)_{+1}=(\nabla R)_{0}=0$ \\
N & $(\nabla R)_{+3}=(\nabla R)_{+2}=(\nabla R)_{+1}=(\nabla R)_{0}=(\nabla R)_{-1}=0$ \\
D & $(\nabla R)_{+3}=(\nabla R)_{+2}=(\nabla R)_{+1}=(\nabla R)_{-1}=(\nabla R)_{-2}=(\nabla R)_{-3}=0$ \\
O & all vanish (symmetric space) \\
\hline
\end{tabular}
}
\caption{The relation between $\nabla$Riemann and the vanishing of boost weight components.}\label{driemtypes}
\end{center}
\end{table} 

In particular, we see that a type D Riemann tensor has $R=(R)_0$. Similarly, a type D 
$\nabla$Riemann tensor has $\nabla R=(\nabla R)_0$. If this holds for all covariant derivatives 
of the Riemann tensor (i.e., $\nabla^{(k)} R=(\nabla^{(k)} R)_0$, for all $k$), then we call the spacetime type D$^k$.

\section{Appendix: Curvature operators and  curvature projectors}
A curvature operator, ${\sf T}$, is a tensor considered as a (pointwise) linear operator 
\[ {\sf T}:~V\mapsto V, \]
for some vector space, $V$, constructed from the Riemann tensor, its covariant derivatives, and the curvature invariants. 
 
The archetypical example of a curvature operator is obtained by 
raising one index of the Ricci tensor.  The Ricci operator is 
consequently a mapping of the tangent space $T_p\mathcal{M}$ into 
itself: 
\[ {\sf R}\equiv(R^{\mu}_{~\nu}):~T_p\mathcal{M}\mapsto T_p\mathcal{M}. \] 
Another example of a curvature operator is the Weyl tensor, considered as an operator, ${\sf C}\equiv (C^{\alpha\beta}_{\phantom{\alpha\beta}\mu\nu}$), mapping bivectors onto bivectors. 
 
For a curvature operator, ${\sf T}$, consider an eigenvector ${\sf 
v}$ with eigenvalue $\lambda$; i.e., ${\sf T}{\sf v}=\lambda{\sf 
v}$. If $d=\mathrm{dim}(V)$ and $n$ is the dimension of the 
spacetime, then the eigenvalues of ${\sf T}$ are $GL(d)$ invariant. Since 
the Lorentz transformations, $O(1,n-1)$, will act via a 
representation $\Gamma\subset GL(d)$ on ${\sf T}$, 
 \emph{the eigenvalue of a curvature operator is an $O(1,n-1)$-invariant curvature scalar}. 
Therefore, curvature operators naturally provide us with a set of 
curvature invariants (not necessarily polynomial invariants) 
corresponding to the set of distinct eigenvalues: $\{\lambda_A 
\}$. Furthermore, the set of eigenvalues are uniquely determined 
by the polynomial invariants of ${\sf T}$ via its characteristic 
equation. The characteristic equation, when solved, gives us the 
set of eigenvalues, and hence these are consequently determined by 
the invariants.

We can now define a number of associated curvature operators. For 
example,  for an eigenvector ${\sf v}_A$ so that ${\sf T}{\sf 
v}_A=\lambda_A{\sf v}_{A}$, we can construct the annihilator 
operator: 
\[ {\sf P}_A\equiv ({\sf T}-\lambda_{A}{\sf 1}).\] 
Considering the Jordan block form of ${\sf T}$, the eigenvalue ${\lambda_A}$ corresponds to a set of Jordan blocks. These blocks are of the form: 
\[ {\sf B}_A=\begin{bmatrix} 
\lambda_A & 0 & 0& \cdots & 0 \\ 
1 & \lambda_A & 0& \ddots  & \vdots \\ 
0      & 1 &\lambda_A& \ddots & 0 \\ 
\vdots & \ddots     &\ddots& \ddots & 0 \\ 
0    &     \hdots &0   &  1      & \lambda_A 
\end{bmatrix}.\] 
There might be several such blocks corresponding to an eigenvalue 
$\lambda_A$; however, they are all such that $({\sf 
B}_A-\lambda_A{\sf 1})$ is nilpotent and hence there exists an 
$n_{A}\in \mathbb{N}$ such that  ${\sf P}_A^{n_A}$ annihilates the 
whole vector space associated with the eigenvalue $\lambda_A$. 
 
This implies that we can define a set of operators $\widetilde{\bot}_A$ with eigenvalues $0$ or $1$ by considering the products 
\[ \prod_{B\neq A}{\sf P}^{n_B}_B=\Lambda_A\widetilde{\bot}_A,\] 
where $\Lambda_A=\prod_{B\neq A}(\lambda_B-\lambda_A)^{n_B}\neq 0$ 
(as long as $\lambda_B\neq \lambda_A$ for all $B$). Furthermore, we can now define
\[ \bot_A\equiv {\sf 1}-\left({\sf 1}-\widetilde{\bot}_A\right)^{n_A}  \]
 where $\bot_A$ 
is a \emph{curvature projector}. The set of all such curvature 
projectors obeys: 
\beq {\sf 1}=\bot_1+\bot_2+\cdots+\bot_A+\cdots, 
\quad \bot_A\bot_B=\delta_{AB}\bot_A. 
\eeq We can use these 
curvature projectors to decompose the operator ${\sf T}$: \beq 
{\sf T}={\sf N}+\sum_A\lambda_A\bot_A. \label{decomp} \eeq The 
operator ${\sf N}$ therefore contains  all the information not 
encapsulated in the eigenvalues $\lambda_A$. From the Jordan form 
we can see that ${\sf N}$ is nilpotent; i.e., there exists an 
$n\in\mathbb{N}$ such that ${\sf N}^n={\sf 0}$. In particular, if 
${\sf N}\neq 0$, then ${\sf N}$ is a negative/positive 
boost weight operator which can be used to lower/raise the 
boost weight of a tensor. 
 
Considering the Ricci operator, or the Weyl operator, we can show 
that (where the type refers to either Ricci type or Weyl type): 
\begin{itemize} 
\item{} Type I: ${\sf N}={\sf 0}$, $\lambda_A\neq 0$. 
\item{} Type D: ${\sf N}={\sf 0}$, $\lambda_A\neq 0$. 
\item{} Type II: ${\sf N}^3={\sf 0}$, $\lambda_A\neq 0$. 
\item{} Type III: ${\sf N}^3={\sf 0}$, $\lambda_A=0$. 
\item{} Type N: ${\sf N}^2={\sf 0}$, $\lambda_A=0$. 
\item{} Type O: ${\sf N}={\sf 0}$, $\lambda_A=0$. 
\end{itemize} 
 
Consider a curvature projector $\bot: T_p\mathcal{M}\mapsto T_p\mathcal{M}$. Then, for a Lorentzian spacetime there are 4 categories: 
\begin{enumerate} 
\item{} Timelike: For all $v^{\mu}\in T_p\mathcal{M}$, $v_{\nu}(\bot)^{\nu}_{~\mu}v^{\mu}\leq 0$. 
\item{} Null: For all $v^{\mu}\in T_p\mathcal{M}$, $v_{\nu}(\bot)^{\nu}_{~\mu}v^{\mu}= 0$. 
\item{} Spacelike: For all $v^{\mu}\in T_p\mathcal{M}$, $v_{\nu}(\bot)^{\nu}_{~\mu}v^{\mu}\geq 0$. 
\item{} None of the above. 
\end{enumerate} 
 
We can consider a complete set of curvature 
projectors, $\bot_A: T_p\mathcal{M}\mapsto T_p\mathcal{M}$, 
which can be of any of the aforementioned categories, and we 
use a Segre-like notation to characterize the set 
with a comma separating time and space. For example, $\{1,111\}$ 
means we have 4 projectors: one timelike, and three spacelike. A 
bracket indicates that the image of the projectors are of 
dimension 2 or higher; e.g., $\{(1,1)11\}$ means that we have two 
spacelike operators, and one with a 2 dimensional image. If there 
is a null projector, we automatically have a second null 
projector. Given an NP frame $\{ 
\ell_{\mu},n_{\mu},m_{\mu},\bar{m}_{\mu}\}$, then a null-projector 
can typically be: 
\[ (\bot_1)^{\mu}_{~\nu}=-\ell^{\mu}n_{\nu}. \] 
Note that $\bot_1^2=\bot_1$, but it is not symmetric. Therefore, acting from the left and right gives two different operators. Indeed, defining 
\[ (\bot_2)^{\mu}_{~\nu}\equiv g_{\nu\alpha}g^{\mu\beta}(\bot_1)^{\alpha}_{~\beta}, \] 
we get a second null-projector being orthogonal to $\bot_1$. The 
existence of null-projectors enables us to pick out certain null 
directions; however, note that the null-operators, with respect to 
the aforementioned NP frame, are of boost weight 
0 and so cannot be used to lower/raise the boost weights. In 
particular, considering the combination $\bot_1+\bot_2$ we see 
that the existence of null-projectors implies the existence of 
projectors of type $\{(1,1)(11)\}$.

\section{Appendix: Previous work}

A number of partial results on 4D $CSI$ spacetimes
can be deduced from previous work. For example, it can be shown that in 4D Petrov (Weyl) type I
and PP- (Ricci) type I
Lorentzian $CSI$ spacetimes are locally homogeneous,
which is consistent with the above analysis.

Consider first the  PP-type I case. In this case, there is a frame in which the Ricci
tensor has components that are all constant and can be put into a canonical
form \cite{CSI3d}, and the result easily follows.

Let us next consider the Petrov type I case. Immediately we know
that local (zeroth order)
curvature homogeneity implies local homogeneity in this case \cite{Mp}. From
\cite{HallBook} it follows that if a 4D spacetime is of
Petrov type I it can be classified according to its rank and it is
either: (i) general curvature class A, or (ii) curvature class C
with restricted Segre type (see \cite{HallBook} or \cite{chpinv}).

Now, suppose the components of the Riemann tensor $R^a \;\!
_{bcd}$ are given in a coordinate domain $U$ with metric $g$. In
case (i), where the curvature class is of type A, for any other
metric $g'$ with the same components $R^a \;\! _{bcd}$ it follows
that $g'_{ab} = \alpha g_{ab}$ (where $\alpha$ is a constant).
We can then pass to the frame formalism and determine the frame
components of the Riemann tensor.  The Petrov
type I case is completely backsolvable \cite{Carminati} and hence
the frame components are completely determined by the zeroth order
scalar invariants, and it follows that the spacetime is
locally homogeneous in this case.

Let us now consider case (ii), where the curvature
class is $C$.  Again, let us suppose that the $R^a \;\! _{bcd}$ are given
in $U$ with metric $g$.  If $g'$ is any other metric with the same
$R^a \;\! _{bcd}$, it follows that
$g'_{ab} = \alpha g_{ab} + \beta k_ak_b$
(where $\alpha$ and $\beta$ are constants).  The equation
\begin{equation} R^a \;\! _{bcd} k^d =  0  \label{star}
\end{equation}
has a unique (up to scaling) non-trivial solution for $k \in T_m M$.
If $R^a \;\! _{bcd;e} k_a \neq 0$, then $\beta =0$ and the metric
is determined up to a constant conformal factor (and the holonomy
type is $R_{15}$
{\footnote{As defined, for example, in \cite{HallBook}.}}).  This is similar to the first case discussed
above (but now some information on the covariant derivative of the
Riemann tensor is necessary;  e.g., $I_2 \equiv [R^{abcd;e}
R_{abcd;e} -4R^{ab;c} R_{ab;c} +R^{,a} R_{,a}] \ne 0$  \cite{chpinv}). Hence, the spacetime is
locally homogeneous.

If $R^a \;\! _{bcd;e} k_a = 0$, then $R^a \;\! _{bcd} k_{a;e} = 0$,
and since eqn. (\ref{star}) has a unique solution, $k_a$ is recurrent.  If $k_a$ is
null, the spacetime is algebraically special, and since we
assume that the Petrov type is I, this is not possible. Hence, $k_a$ is (a)
timelike (TL) or (b) spacelike (SL) and is, in fact,
covariantly constant (CC).
In case (iia), the spacetime admits a TL CC vector field $k_a$.
The holonomy is $R_{13}$, with a TL holonomy invariant subspace
which is non-degenerately reducible, and $M$ is consequently
locally $(1+3)$ decomposable (and static). All of the non-trivial
components of the Riemann
tensor and its covariant derivatives are constructed from the $3D$
positive definite metric, and if the spacetime is $CSI$ it is consequently
locally homogeneous.
In case (iib), the spacetime admits a SL CC vector field $k_a$.
The holonomy is $R_{10}$, there exists a holonomy invariant SL
vector $k_a$ which is non-degenerately reducible, and $M$ is this
locally $(3+1)$ decomposable.
Classification now reduces to the classification of a subclass of
$3D$ Lorentzian spacetimes, and it follows from \cite{CSI}
that the spacetime is locally homogeneous.

The analysis could proceed in a similar fashion on a case-by-case
basis (according to Petrov and/or Segre type); however, it is not
clear a complete proof could be obtained in this way. In addition, the analysis
presented above in the main text of the paper is easier to apply and is more readily
applicable to higher dimensional generalizations.

{\em Acknowledgements}. We would like to thank Robert Milson for helpful comments on the current manuscript.
This work was supported by NSERC of Canada.

\end{document}